# Spin-entropy contribution to thermopower in the $[Ca_2CoO_{3-t}]_{0.62}(CoO_2)$ misfits

J. Hejtmánek, Z. Jirák, and J. Šebek

*Institute of Physics of ASCR, Na Slovance 2, 182 21 Praha 8, Czech Republic*

**Abstract**

Two samples of the $[Ca_2CoO_{3-t}]_{0.62}(CoO_2)$ misfit cobaltate, often denoted as the $Ca_3Co_4O_9$ phase, were prepared from the same ceramic material by the oxygen and argon annealing, resulting in different carrier concentrations in the conducting $CoO_2$ layers, $n$=0.31 and 0.19 hole/Co, respectively. Electrical and thermal transport properties were studied in dependence of magnetic field up to 140 kOe. The magnetothermopower data reveal an extra spin-entropy contribution to Seebeck coefficient that is not expected for carriers of Fermi liquid character. Its magnitude is unprecedentedly large and makes at zero field up to 50% of the theoretical limit $\frac{k_B}{|q_e|}\ln 2$ = 59 µV/K. This spin-entropy contribution is gradually suppressed with increasing magnetic field, and the saturation is even observed when temperatures are low enough. To understand the results, the thermopower is treated in terms of purely thermodynamic Kelvin formula, and so-called Spin liquid model is evoked, providing a reason for the spin-entropy manifestation in the $[Ca_2CoO_{3-t}]_{0.62}(CoO_2)$ misfits.

**Introduction**

The layered cobaltates of the $Na_xCoO_2$ or so-called misfit types are generally considered as promising systems for thermoelectric applications with respect to their high Seebeck coefficient, good electrical conductivity, low toxicity and chemical stability at elevated temperatures [1, 2]. As the crystal structure is concerned, these systems have in common hexagonal layers of edge shared $CoO_6$ octahedra, in which electrical transport is carried out. These $CoO_2$ layers alternate with planes of variable content of $Na^+$ ions or, in the "misfits", there are interpolating blocks of two-, three-, or four-plane oxide layers of rock-salt (RS) type. In particular, the commonly known misfit system of approximate stoichiometry $Ca_3Co_4O_9$ contains a block formed by the CoO plane sandwiched by two CaO planes. The lattice parameters of the $CoO_2$ layers and $Ca_2CoO_3$ triple RS blocks are incommensurate in the basal $b$-direction. In the block compositions, the appropriate chemical formula at full oxygen content is $[Ca_2CoO_3]_{0.62}(CoO_2)$.

It is well known that $Na_xCoO_2$ with full sodium content x=1 is a band insulator with $Co^{3+}$ ions in the low-spin (LS) state, whereas sodium deficient systems with cobalt ions in the mixed LS $Co^{3+}/Co^{4+}$ valence show a rich spectrum of behaviours - see e. g. [3]. With respect to the applications, the main interest is given to compositions close to x≈0.70 (the doping concentration $n$=0.30 hole per $CoO_2$), which are metallic conductors with Fermi liquid characteristics at the lowest temperatures [4] and, importantly, they show unusual steepness of quasilinear dependence of Seebeck coefficient at low temperatures and permanent increase at room and elevated temperatures up to very high positive values ($S \approx$ 100 -170 $\mu$V K$^{-1}$ in the range 300 - 1000 K) [1, 5]. As the present misfits is concerned, the $Ca_2CoO_{3-t}$ triple RS blocks are supposed to be always in $Co^{3+}$ state, either of the intermediate-spin (IS) or high-spin (HS) character [6, 7, 8]. In that case the hole concentration in conducting $CoO_2$ layers is nominally $n$=0.38, but it is decreased to about $n$=0.32 hole per $CoO_2$ due to some oxygen deficiency in RS blocks at standard sample preparation, typically $t$≈0.05. In distinction to sodium cobaltates there is a clear localization in the charge transport in $CoO_2$ layers, possibly due to random potential induced by charge and spin arrangements in the adjacent triple RS blocks. Nevertheless, Seebeck coefficient retains the steeply increasing "metallic-like" dependence and reaches similarly large values as found in comparable $Na_xCoO_2$ systems.

Within Boltzmann transport theory, the diffusion Seebeck coefficient is given by a ratio of two Fermi integrals

$$S_{Diff} = \frac{k_B}{q_e} \int_0^W N(\varepsilon) v_x(\varepsilon)^2 \tau(\varepsilon) \frac{(\varepsilon-\mu)}{k_B T}\left(-\frac{\partial f(\varepsilon)}{\partial \varepsilon}\right) d(\varepsilon) / \int_0^W N(\varepsilon) v_x(\varepsilon)^2 \tau(\varepsilon) \left(-\frac{\partial f(\varepsilon)}{\partial \varepsilon}\right) d(\varepsilon), \quad (1)$$

the limiting form of which is the well-known $T$-linear Mott thermopower for metals. Here $T$ is the absolute temperature, $q_e$ is the electron charge, $v_x$ is the projection of group velocity into the transport direction and $\tau$ is the relaxation time that is often supposed to be energy independent at vicinity of chemical potential $\mu$. It is seen that Seebeck coefficient is related to some average value of the energy difference with respect to chemical potential $(\varepsilon - \mu)$. Since this value is multiplied by the derivation of Fermi-Dirac distribution function $-\frac{\partial f(\varepsilon)}{\partial \varepsilon}$, the main weight in above mentioned integration have the more-than-half populated states (hole carriers) of energy $\mu-k_BT$ and the less-than-half populated states (electron carriers) of energy $\mu+k_BT$ that give a positive and negative contribution, respectively, and nearly cancel in common metals. Very large Seebeck coefficient is possible only when the respective density of states or electron group velocities are very unlike below and above the chemical potential, which is indeed assumed to occur in the present cobaltates due to a particular "pudding mold" character of the $a_{1g}$ subband in which Fermi level is located [9]. The numerical calculations performed within the generalized Mott formula (1) with a use of the *ab-initio* determined electronic structures for $Na_{0.7}CoO_2$ provide, indeed, a large positive Seebeck coefficient with temperature dependence close to what is actually observed [10].

Nonetheless, in spite of apparent success of the generalized Mott formula, there are two contradictory points. First one is the short mean free path of carriers that classifies layered cobaltates as bad metals. In such a case, the purely thermodynamic Kelvin formula $S_{Kelvin} = \frac{1}{q_e}\left(\frac{\partial \mu}{\partial T}\right)_{n,V} = \frac{1}{q_e}\left(\frac{\partial S}{\partial n}\right)_{T,V}$ is considered to be more appropriate [11, 12]. Most importantly the Kelvin formula: (i) defines Seebeck coefficient simply as "entropy per carrier", divided by the carrier charge and (ii) when applied to continuous spectrum of fermion energies, an expression omitting eventual role of the particle diffusion lengths but otherwise similar to formula (1) is obtained:

$$S_{Kelvin} = \frac{k_B}{q_e} \int_0^W N(\varepsilon) \frac{(\varepsilon - \mu)}{k_B T}\left(-\frac{\partial f(\varepsilon)}{\partial \varepsilon}\right) d(\varepsilon) / \int_0^W N(\varepsilon)\left(-\frac{\partial f(\varepsilon)}{\partial \varepsilon}\right) d(\varepsilon) \qquad (2)$$

(see Appendix A for more details). The second questionable point refers to magneto-transport phenomena, especially the large suppression of Seebeck coefficient that scales with magnetic field following the entropy formula for independent spins. This effect was reported firstly for $Na_{0.68}CoO_2$ and later on in for the $[Bi_{1.7}Co_{0.3}Ca_2O_4]_{0.60}(CoO_2)$ misfit with quadruple RS blocks [13, 14]. In this paper, we report on similar experiments on two $[Ca_2CoO_{3-t}]_{0.62}(CoO_2)$ samples that have been prepared from the same ceramic material by $O_2$ and Ar annealing, and possessed very different doping levels ($n=0.31$ and $0.19$ hole per $CoO_2$, respectively). In both samples, we observe a suppression of Seebeck coefficient in high external field, which reaches up to 50% of the theoretical value derived for independent spins, $\frac{k_B}{|q_e|}\ln 2 = 59$ μV/K. Such huge effect can be hardly explained by a common splitting of spin-majority and spin-minority densities of states in available external magnetic fields [15], and points rather to an existence of singly occupied carrier states with spin degree of freedom. Such carriers seem to compete with quasiparticle states of Fermi liquid character, which are characterized by *k*-vectors and allow a double occupation (spin-up and spin-down). In this context, we would like to turn attention to so-far hypothetical model of Spin liquid, introduced by Spałek and his collaborators [16, 17, 18], in which *the double occupation is eliminated in principle*.

As shown in detail in Appendix B, the Kelvin formula yields for the Spin-liquid particles a sum of two terms. The first one can be formally interpreted as thermopower of (*non-interacting*) spinless fermions depending on the energy spectrum of the particles. It can be expressed through density of available states according to the general formula mentioned above for common fermions - see expression (2). The second term within the Spin-liquid approximation is given by expression:

$$S_{Kelvin,2} = \frac{k_B}{|q_e|}\left[\ln(2\cosh(\mu_B H/k_B T)) - (\mu_B H/k_B T)\tanh(\mu_B H/k_B T)\right], \quad (3)$$

or $S_{Kelvin,2} = \frac{k_B}{|q_e|}\ln 2 = 59\ \mu V/K$ for $H=0$. This term can be interpreted as the spin-entropy contribution of (*non-interacting*) bare spins ½. The Spin-liquid model thus bears, at least qualitatively, the features we observe on $[Ca_2CoO_{3-t}]_{0.62}(CoO_2)$ samples at low temperatures – the steeply increasing "metallic-like" thermopower (2), complemented with a large spin entropy term (3) that can be suppressed by magnetic field at low temperatures.

For a deeper understanding of the particles carrying spin entropy in unprecedented amount, we revisit also our previously published heat capacity data [19]. In the present very detailed and complex analysis, the $CoO_2$ layers in studied misfits emerge as electronic systems at the crossover between the metallic and hopping regimes. We propose that the enigmatic spin entropy contribution to thermopower arises namely due to a dynamic equilibrium between the itinerant and more-or-less localized carrier states.

## I. Experimental part

The $[Ca_2CoO_{3-t}]_{0.62}(CoO_2)$ phase was prepared by ceramic route through the calcinations of oxides and carbonates with several intermittent grindings. The sintering of pressed tablet resulted in slightly porous material with actual density 3.84 gcm$^{-3}$, corresponding to 82% of the X-ray derived value. One piece cut into a parallelepiped form of dimensions 10x4x2 mm$^3$ was annealed in oxygen, firstly for 48 hours at 830°C followed by 48 hours at 730°C and finally 240 hours at 630°C. The final product, denoted as sample A, showed high "metallic-like" conductivity both at low and high temperatures. Based on a comparison with the literature data for $[Ca_2CoO_{3-t}]_{0.62}(CoO_2)$ ceramics, where carefully determined oxygen content and electrical properties were correlated [20], the oxygen deficiency in RS blocks of sample A was set as $t \approx 0.06$ (the doping concentration $n=0.31$ hole per $CoO_2$). The second specimen was fabricated from the same as-prepared $[Ca_2CoO_{3-t}]_{0.62}(CoO_2)$ material by annealing in argon at 630°C for 36 hours. This yielded the oxygen deficient sample B, for which the value $t \approx 0.15$ (the doping concentration $n=0.19$ hole per $CoO_2$) was estimated based on similar criteria as mentioned above (taking into account mainly the significant electronic localization at low temperatures as seen in the inset of Fig. 3). Because of the composite structure of the misfits and our experimental limitations, we were not able to assess the hole concentration in $CoO_2$ layers in a more direct way, but the very different doping levels of our samples became evident from their different electrical properties and different volume characteristics like the magnetization data, or as shown in Ref.[19], from different specific heat data. Although some uncertainty may remain, we do believe that the hole concentrations for the O2-annealed sample A (~ 0.31 hole/$CoO_2$) and the Ar-annealed sample (~ 0.19 hole/$CoO_2$) represent a credible estimate.

Basic magnetic characterization of samples A and B was done on a SQUID magnetometer (MPMS-XL;Quantum Design) working in the range 2–400 K. The zero-field data on the electrical conductivity, thermopower and thermal conductivity were taken simultaneously in the temperature range 3.5 -310 K using a home-made cell attached to the close-cycle cryostat. The four-contact method was applied for the electrical and thermal transport measurements, the contact were realized using silver paint and silver wires. Details of the measurement can be found elsewhere [21]. The magnetotransport experiments at high fields up to 140 kOe were performed at 2-300 K using a commercial TTO option of Quantum Design PPMS device. Most importantly, we noticed a good agreement (relative error ≤ 5%) between the zero-field thermal transport data acquired either via home-made or using PPMS TTO, the fact that supports reliability of the measured data.

The heat capacity data discussed in Part III were taken from our previous study performed on a similar $[Ca_2CoO_{3-t}]_{0.62}(CoO_2)$ couple [19]. They concern namely the sample prepared through classical ceramic route (CCR) by sintering of cold pressed powder in oxygen, and the spark plasma sintered sample (SPS) at 900°C in vacuum. The rather porous sample CCR (relative density of 77%) was found to be highly oxygenized with $t \approx 0.04$, corresponding to the doping concentration $n=0.33$ hole per $CoO_2$, whereas the dense sample SPS (relative density of 98%) was oxygen deficient with $t \approx 0.17$, corresponding to the doping concentration $n=0.17$ hole per $CoO_2$.

## II.  Results

The magnetic properties of the $[Ca_2CoO_{3-t}]_{0.62}(CoO_2)$ samples A (O$_2$ annealed) and B (Ar annealed) are manifested by the field-cooled (FC) susceptibility in Fig. 1 and the low-temperature hysteresis loops in Fig. 2. The results are complemented by data for the utmost oxygenized ceramics CCR studied previously [19]. It is seen that CCR ($n$=0.33 hole per $CoO_2$) and the present sample A ($n$=0.31 hole per $CoO_2$) are characterized by a steep susceptibility increase below 20 K, which signals the onset of long- or short-range magnetic ordering. These two oxygenized samples exhibit indeed a finite remanence, although its absolute value is significantly lower for sample A as seen in hysteresis loops in Fig. 2. The temperature at which the remanence vanishes on heating at zero field is determined to $T_c \approx$ 12 K and 16 K for sample A and CCR, respectively (see the remanence data in the inset of Fig. 1). On the other hand, both the remanence data and hysteresis loop recorded for the Ar-annealed sample B ($n$=0.19 hole per $CoO_2$) point to almost purely "paramagnetic" behaviour without any significant tendency to a long-range order down to 2 K. We may thus conclude that the hole carriers in the $CoO_2$ layers experience non-zero molecular field $H_m$ of Weiss type not only in sample CCR but also in A, whereas the existence of such molecular field in sample B is questionable.

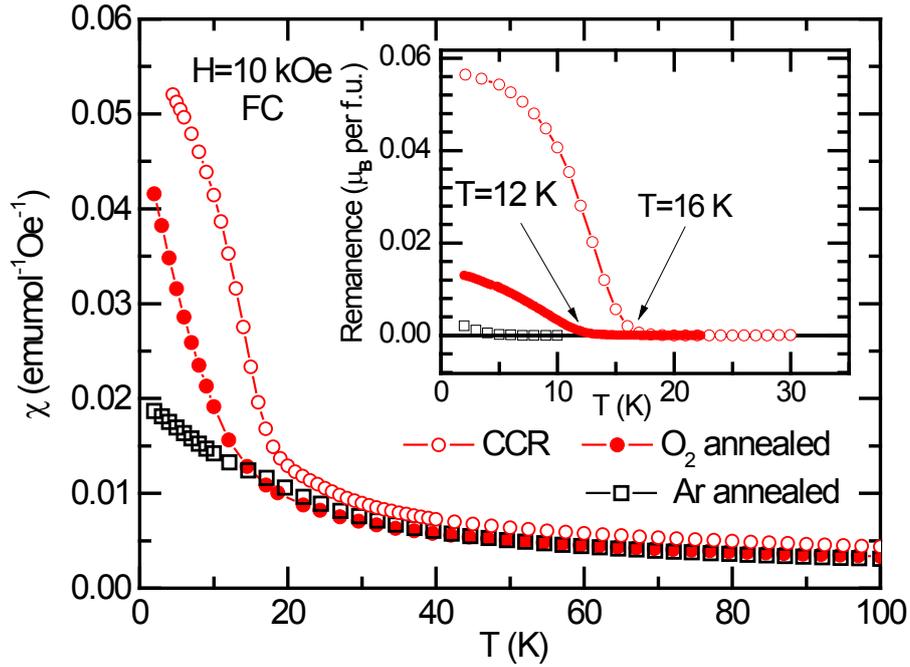

Fig. 1. (Color online) The magnetic susceptibility at 10 kOe (measured in field cooled mode) for the oxygenized $[Ca_2CoO_{3-t}]_{0.62}(CoO_2)$ sample A ($t≈0.06$) and the reduced sample B ($t≈0.15$). Data for the previously studied ceramics CCR ($t≈0.04$) are included for comparison with sample A. The inset shows the remanence.

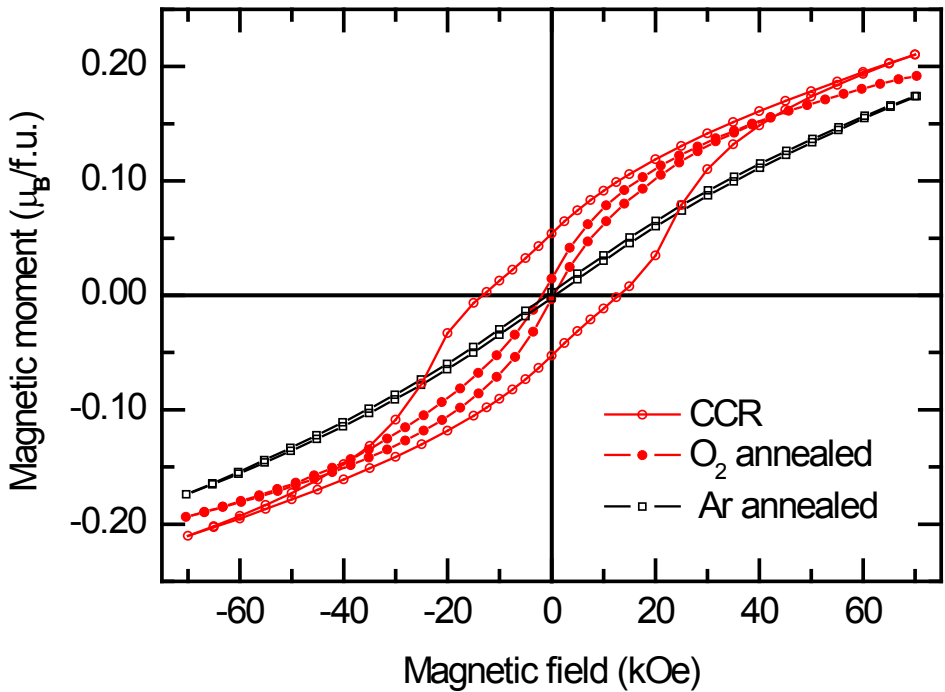

Fig. 2. (Color online) The hysteresis loops taken at 2 K for the $[Ca_2CoO_{3-t}]_{0.62}(CoO_2)$ samples A, B, and CCR. Note the marked difference in remanent moment and coercitive field for the oxygenized samples A ($t≈0.04$) and CCR ($t≈0.06$).

The magnetization curves at 2 K (Fig. 2) are suggesting for coexistence of two components. First, one may note the high-field magnetization para-process similar for all studied samples. We propose that this para-process likely reflects the response to external field of the AFM configuration of IS or HS $Co^{3+}$ moments in RS layers. The second component can be ascribed to the spins of itinerant and localized holes in $CoO_2$ layers. Their contribution is manifested by the steep magnetization rise in low and medium fields, accompanied with more-or-less marked hysteresis in the oxygenized samples CCR and A. The fact that the $CoO_2$ related component tends quickly to a saturation while the para-process survives to very high fields is likely the reason for the continuing magnetoresistance when magnetothermopower is already saturated, as seen in Figs. 5, 6 and 8, 9 below.

The electrical resistivity of the $O_2$-annealed sample A and Ar-annealed sample B is presented in Fig. 3. The results show features that are generally reported for $[Ca_2CoO_{3-t}]_{0.62}(CoO_2)$ misfits, namely the existence of resistivity minimum and the charge carrier localization at the lowest temperature, see also [6]. As shown in the inset of Fig. 3, the localization is particularly strong for the oxygen deficient sample B, and follows an $\exp(AT^{-1/4})$ dependence, which commonly signals the regime of variable range hopping (VRH). The resistivity of the oxygenized sample A has a similar VRH reminding trend at low temperatures, but much weaker. It can be noted that the characteristic activation energy for this sample is well below the thermal energy limit of $k_BT$.

In the Fig. 4 we demonstrate the temperature dependence of the thermal conductivity of samples A and B. Taking into account the low absolute value of the electrical resistivity and applying the Wiedemann-Franz law, the itinerant carriers make only negligible contribution to the displayed data. In this respect the main part of thermal transport is mediated by lattice dynamics.

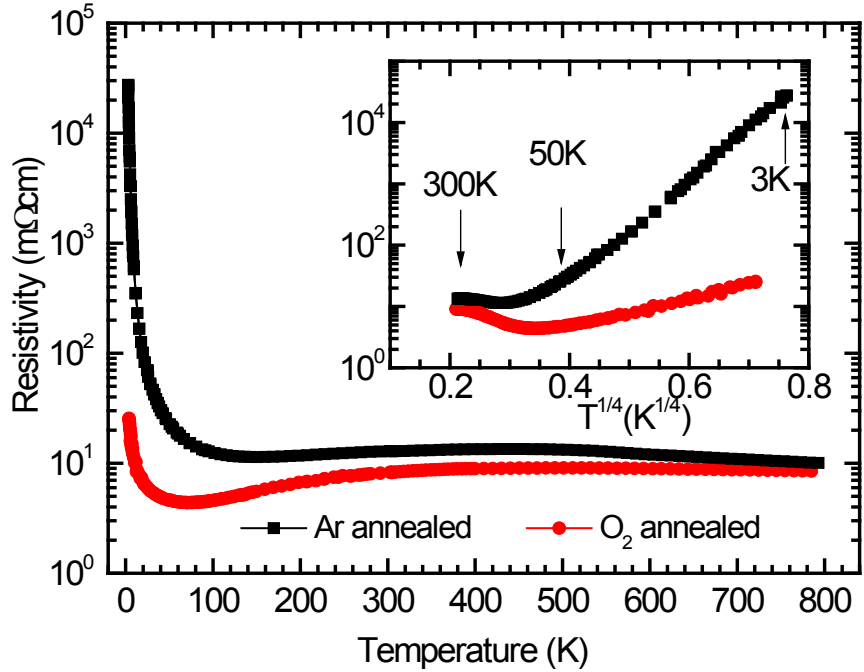

Fig. 3. (Color online) The electrical resistivity for the oxygenized $[Ca_2CoO_{3-t}]_{0.62}(CoO_2)$ sample A ($t \approx 0.06$, the doping $n=0.31$ hole per $CoO_2$) and the reduced sample B ($t \approx 0.15$, $n=0.19$ hole per $CoO_2$). The inset shows the plot of $\log(\rho)$ vs. $T^{-1/4}$ with linear low-temperature trend demonstrating the viability of the VRH scenario.

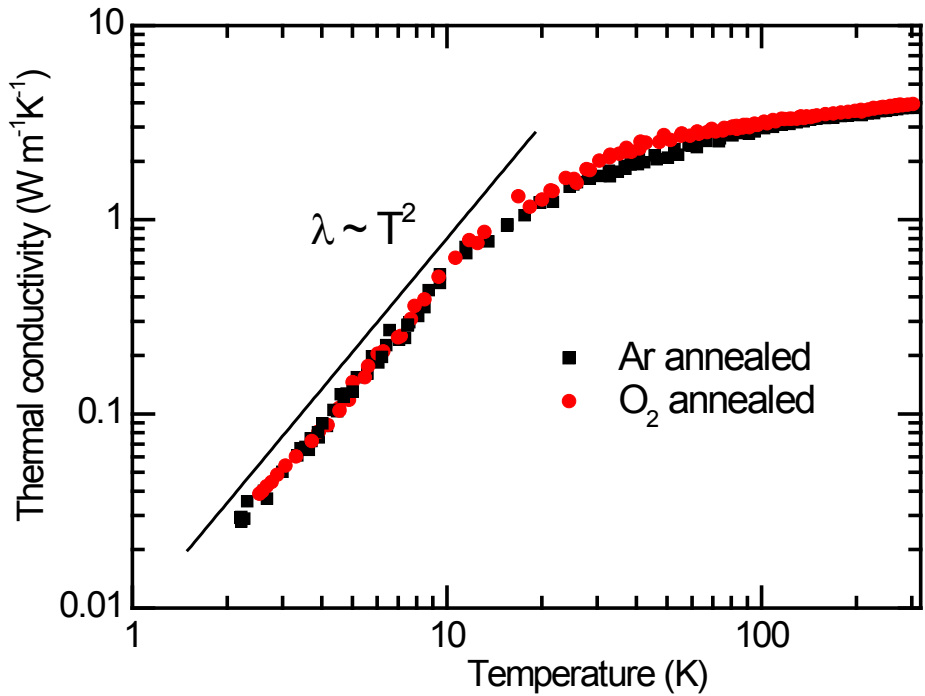

Fig. 4. (Color online) Temperature dependence of the thermal conductivity for the oxygenized $[Ca_2CoO_{3-t}]_{0.62}(CoO_2)$ sample A ($t \approx 0.06$) and the reduced sample B ($t \approx 0.15$).

The low-temperature ($T$<10 K) behaviour, following roughly $T^2$ dependence - see Fig. 4, reflects likely that phonons are predominantly scattered by defects other than grain boundaries, otherwise the temperature dependence would follow the $T^3$ course. Tentatively, we associate the observed quadratic dependence to interaction of long wave-length phonons with scattering centres like sheet-like faults and/or dislocations, the defects that are naturally anticipated in the case of incommensurate layered systems.

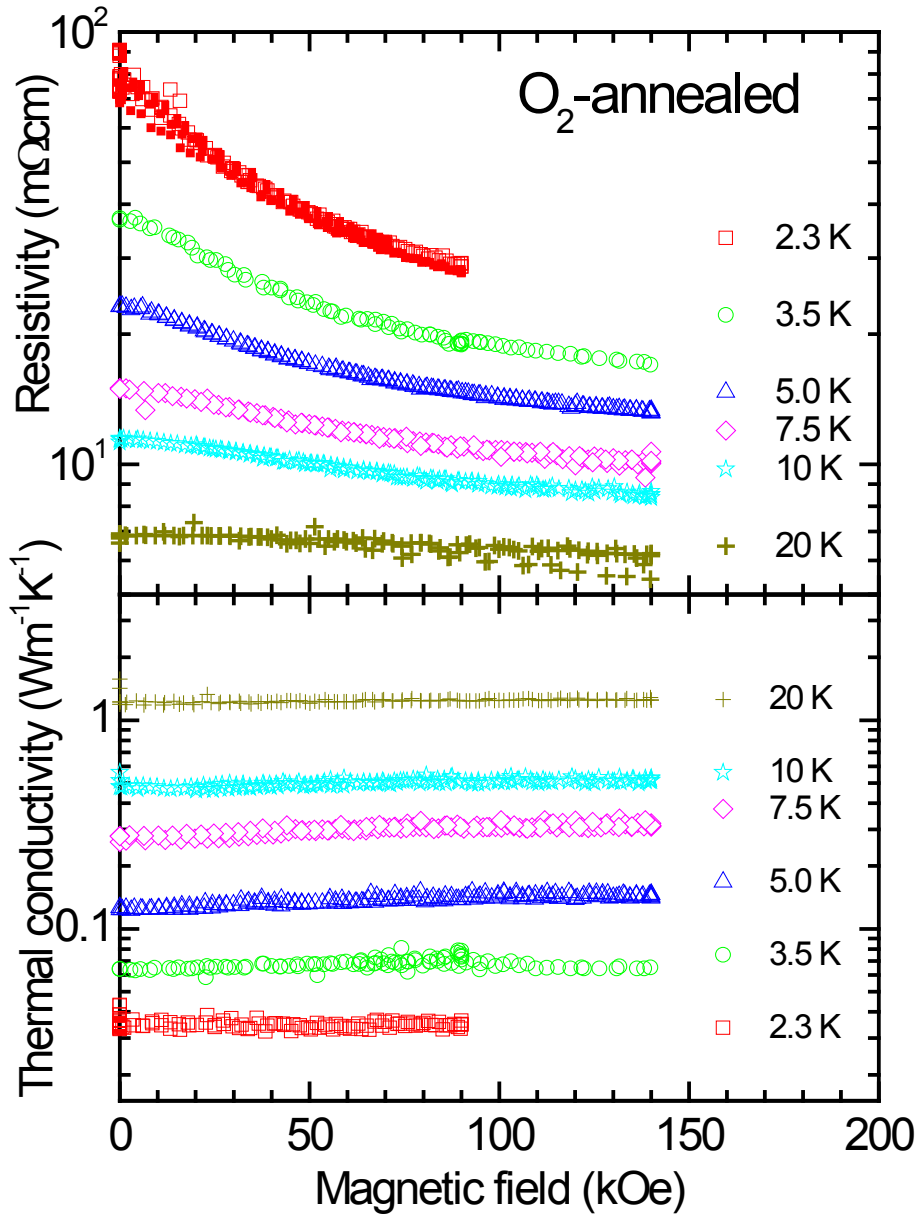

Fig. 5. (Color online) The magnetic field dependence of the electric and thermal transport in the oxygenized [$Ca_2CoO_{3-t}$]$_{0.62}$($CoO_2$) sample A ($t\approx 0.06$).

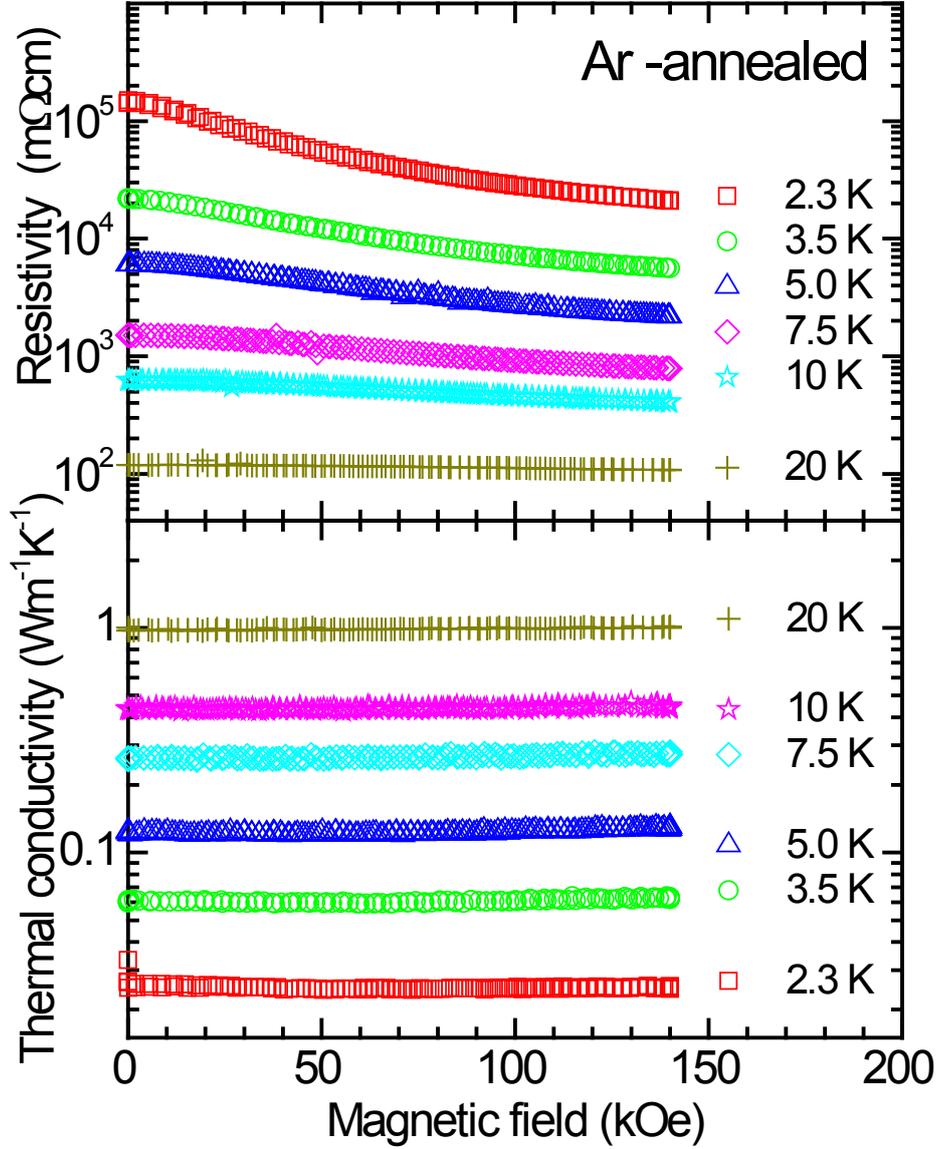

Fig. 6. (Color online) The magnetic field dependence of the electric and thermal transport in the reduced $[Ca_2CoO_{3-t}]_{0.62}(CoO_2)$ sample B ($t \approx 0.15$).

The response of the electric and thermal conductivities to external magnetic field is documented for the $O_2$-annealed sample A and Ar-annealed sample B in Figs. 5 and 6, respectively. These viewgraphs evidence a large negative magnetoresistance, whereas the thermal conduction is evidently not affected. The magnetoresistance effect increases with decreasing temperature, and in spite of very high magnetic field of 140 kOe, no saturation is reached even at 2 K. This can be explained considering that quasiparticle carriers in $CoO_2$ layers continue to be scattered by local potential fluctuations, which are imposed by the AFM arrangements of IS or HS $Co^{3+}$ moments in adjacent RS layers. As the thermal conductivity is concerned, this is field independent down to 2 K up to the fields of 140 kOe. The absence of observable magneto-thermal conductivity effect suggests that neither the itinerant carriers nor the spin excitations play any significant role in the thermal transport. These facts confirm our conclusion that the glass-like thermal conductivity, almost identical for both the $O_2$ and Ar annealed samples, is essentially controlled by the lattice dynamics and phonon scattering on characteristic lattice and morphology defects.

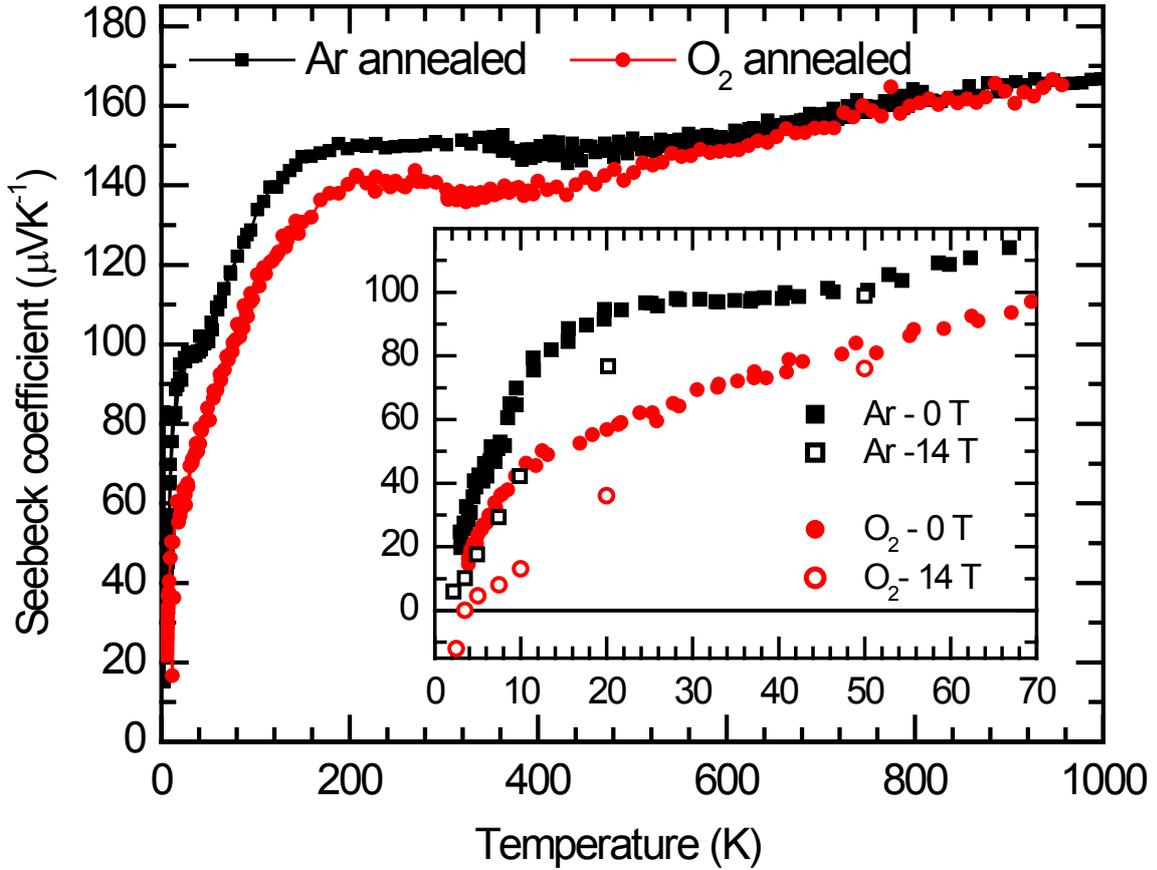

Fig. 7. (Color online) Seebeck coefficient for the oxygenized $[Ca_2CoO_{3-t}]_{0.62}(CoO_2)$ sample A ($t \approx 0.06$, the doping $n=0.31$ hole per $CoO_2$) and the reduced sample B ($t \approx 0.15$, $n=0.19$ hole per $CoO_2$). The inset shows the low-temperature detail together with the measurements at 140 kOe for selected temperatures.

The thermopower data for samples A and B measured at zero field (see Fig. 7) show characteristics commonly known for misfits of the $[Ca_2CoO_{3-t}]_{0.62}(CoO_2)$ type. It is seen that Seebeck coefficient at room and elevated temperatures is high (135 and 150 μV/K for A and B, respectively) and nearly temperature independent up to 400 K. Here the very small anomaly related to some structural disordering in RS layers at ≈ 400 K [22] is not perceptible regarding the used scale. Above this point, a slow linear increase of Seebeck coefficient is found. The behaviour at low temperatures is also gradual. Below the "plateau" down to ≈150 K the thermopower exhibits a sudden change to nearly linear decline that seems to extrapolate to a finite values of ≈40 and 60 μV/K at zero temperature for samples A and B, respectively. The detailed inspection of low temperature behaviour unveils that, as shown in detail in the inset of Fig. 7, the thermopower below 50 K either monotonously decreases to zero (sample A) or adopts within temperature window 20-50 K a constant value of ≈100 μV/K followed by sharp decrease below 10 K.

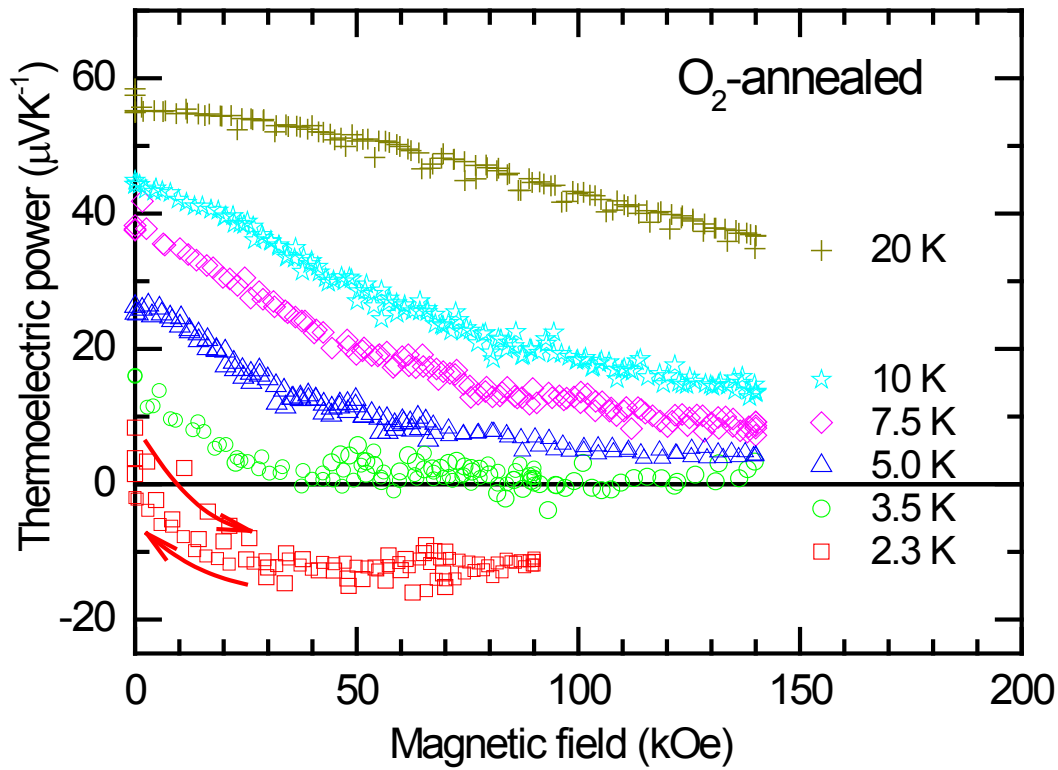

Fig. 8. (Color online) The magnetic field dependence of Seebeck coefficient in the oxygenized $[Ca_2CoO_{3-t}]_{0.62}(CoO_2)$ sample A ($t\approx0.06$).

The thermopower data taken at constant temperatures as a function of applied magnetic field are presented in Figs. 8 and 9. There is a saturation of the field effects on Seebeck coefficient when temperatures are low enough (T≤10K) and a tendency to saturation at higher fields is seen also at higher temperatures up to about 20K. As to our knowledge, the observed suppression of thermopower by external magnetic field is unprecedentedly large in both samples. It reaches up to $\Delta S \approx 30$ μV/K at 140 kOe for T=10 and 7.5 K and, even at the lowest temperature of T ≈ 2 K, the field induced suppression of thermoelectric power makes roughly $\Delta S$ = 15-20 μV/K for both samples. The insensitivity of the negative magnetothermopower to different doping levels points to the common and robust physical background of the observed effect, which we naturally associate, taking into account the evidently "localized" feature of charge carrier transport at low temperatures, with the magnetic field induced quenching of the carrier spin entropy. The quantitative assessment that comes out from experimentally determined suppression of $\Delta S \approx 30$ µV/K confronted with the theoretical limit $\frac{k_B}{|q_e|}\ln 2 = 59$ µV/K is discussed in detail hereafter. Here we only underline the concluding deduction that the spin-entropy carrying holes represent effectively ≈50% of total doping, independently on the actual doping level in $[Ca_2CoO_{3-t}]_{0.62}(CoO_2)$.

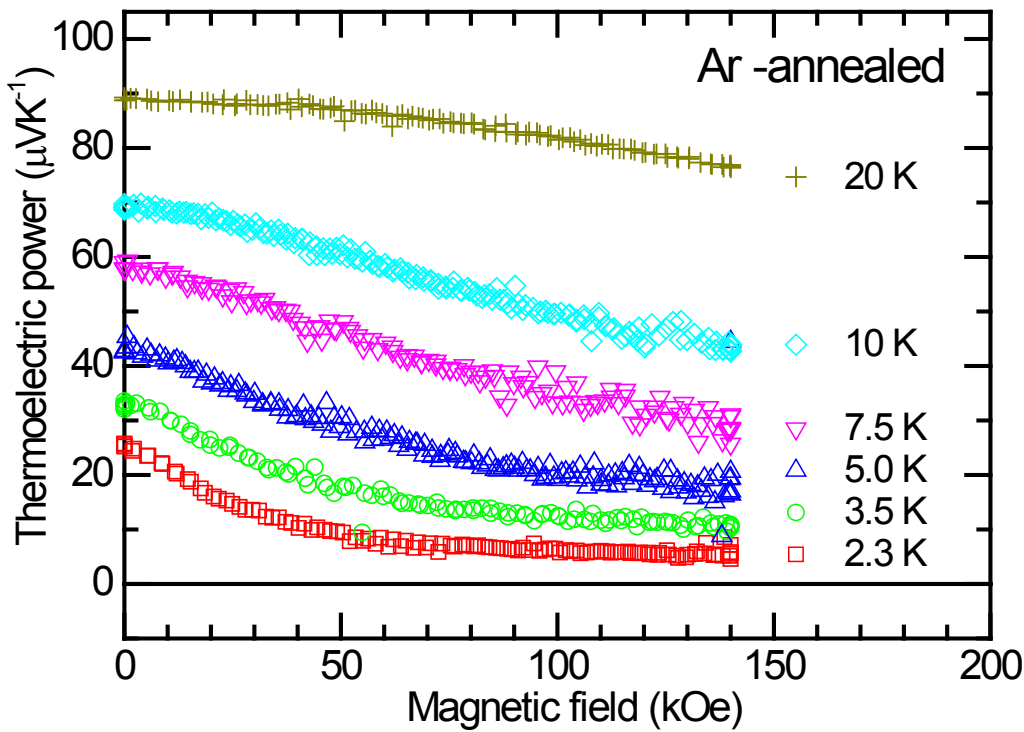

Fig. 9. (Color online) The magnetic field dependence of Seebeck coefficient in the reduced $[Ca_2CoO_{3-t}]_{0.62}(CoO_2)$ sample B ($t\approx 0.15$).

## IV. Discussion

### *Specific heat*

No heat capacity experiments were performed on the present $[Ca_2CoO_{3-t}]_{0.62}(CoO_2)$ samples A and B, but we can make use of the data on another couple of samples differing similarly in the charge doping, namely the highly oxygenized ceramics (CCR) and the oxygen deficient spark-plasma-sintered specimen (SPS) studied in Ref.[19]. The aim is to demonstrate an extreme complexity of charge carriers and spins in the [Ca$_2$CoO$_{3-t}$]$_{0.62}$(CoO$_2$) misfits, namely the broad spectrum of localization effects and characteristic times.

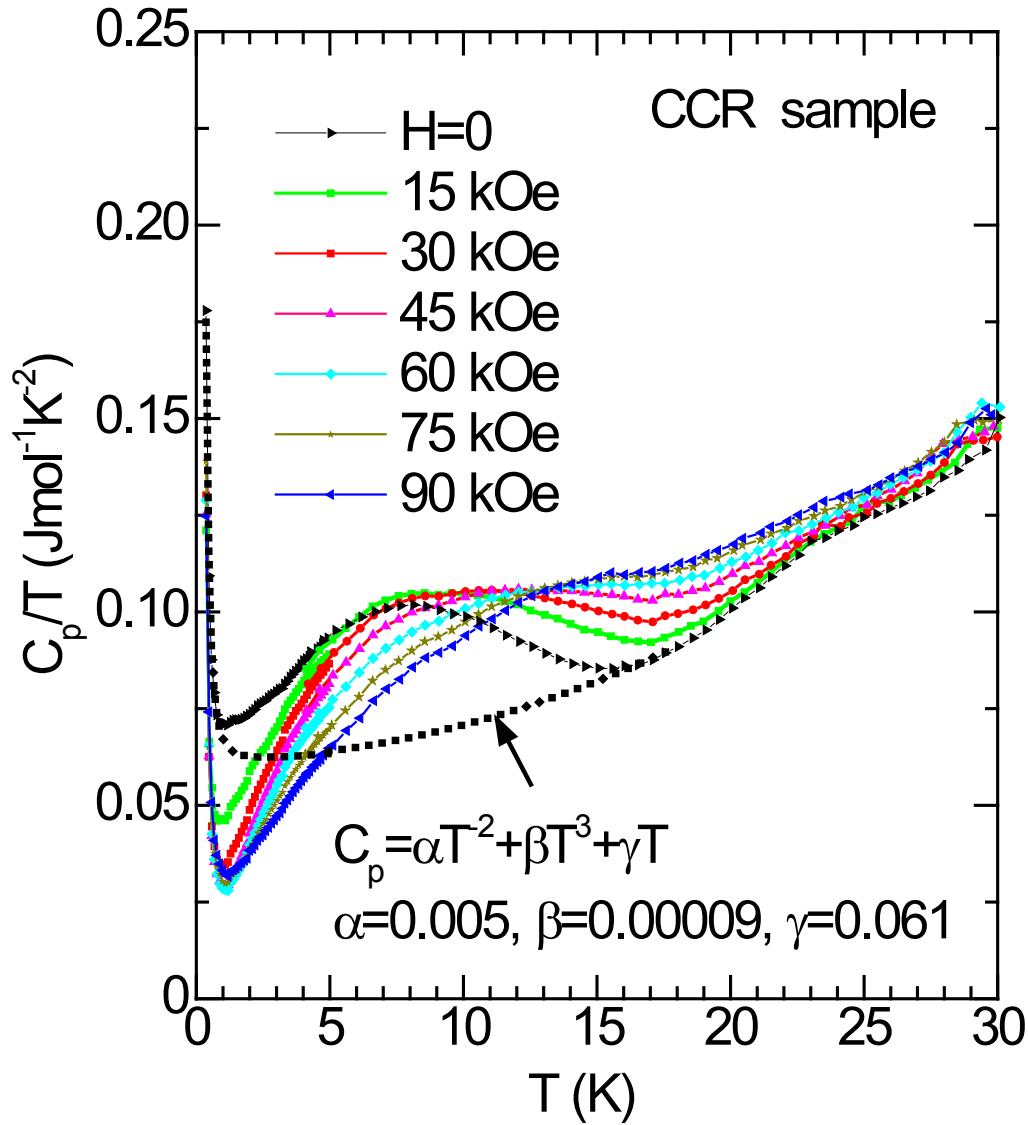

Fig. 10. (Color online) Heat capacity of the [Ca$_2$CoO$_{3-t}$]$_{0.62}$(CoO$_2$) ceramic sample of Ref. [19] ($t \approx 0.04$, the doping $n=0.33$ hole per CoO$_2$), plotted as $C_p/T$ for selected fields in the 0-90 kOe range. The dotted line attached to the zero field data shows a sum of common contributions to heat capacity, namely the nuclear term $\alpha T^{-2}$, linear term $\gamma T$ and lattice phononic term $\beta T^3$ that mark together a baseline for the LS Co$^{4+}$ related Schottky peak that is centered at 7.7 K in the $C_p/T$ data.

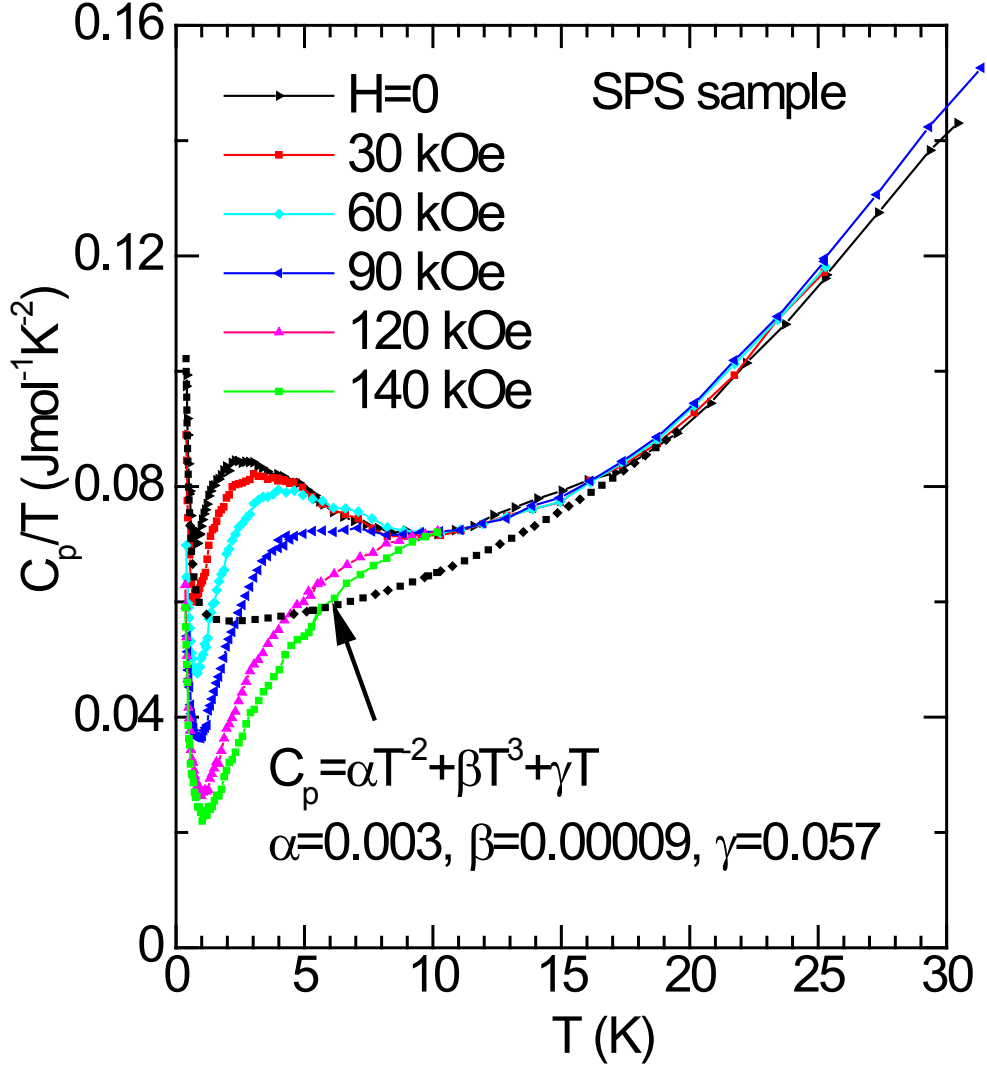

Fig. 11. (Color online) Heat capacity of the spark plasma sintered $[Ca_2CoO_{3-t}]_{0.62}(CoO_2)$ of Ref. [19] ($t≈0.17$, $n=0.17$ hole per $CoO_2$), plotted as $C_p/T$ for selected fields in the 0-140 kOe range. The LS $Co^{4+}$ related Schottky peak is centered at 2.3 K.

The data reproduced in Figs. 10 and 11 show several contributions, among which we note first the very large linear term $\gamma T$ of heat capacity with coefficient $\gamma ≈ 60$ mJmol$^{-1}$K$^{-2}$ for both samples. Under increasing external field, the $\gamma$ value is gradually reduced as can be readily seen at about $≈1$ K where nuclear and lattice contributions to $C_p/T$ are negligible. One may infer from the CCR data in Fig. 10 that the $\gamma$ value at 1 K drops steeply from the very beginning ($H_{ext}$=0-15 kOe) and the saturation is achieved for 75-90 kOe. For the SPS data in Fig. 11 the notable reduction of $\gamma$ shifts to higher fields and the saturation is approached only at 140 kOe. (The differences between these two samples are evidently in a correlation with different trends of their magnetization curves.)

We interpret the residual part $\gamma_e$ as a contribution of fermions in $CoO_2$ layers, having a continuous spectrum with finite density of states at Fermi level. Let us note that this residual value, making for both samples $\gamma_e \approx 20$ mJK$^{-2}$ per $CoO_2$, is still two or three times higher than the value derived from the *ab-initio* electronic calculations, which points to the strongly correlated character of conducting fermions [23].

Another notable feature is the increase of $C_p/T$ data below 1 K, which has a form of common Schottky anomaly $\alpha T^{-2}$ and can be ascribed to $^{59}$Co nuclear spins. This nuclear term does not show any observable change under applied field, but differs for our two samples - $\alpha = 0.005$ and $0.003$ Jmol$^{-1}$K for sample CCR and SPS, respectively, from which the average hyperfine field acting on the rock-salt $^{59}$Co nuclei can be deduced as $H_{hf}$= 290 and 210 kOe. These rather high hyperfine fields and mentioned insensitivity to external fields allow us to conclude that the observed $\alpha T^{-2}$ term is a contribution of $^{59}$Co nuclear spins belonging to the IS/HS Co$^{3+}$ ions in RS layers. Namely, the present values are comparable to the hyperfine fields observed on FM perovskite-type systems La$_{1-x}$Sr$_x$CoO$_3$ [24]. The lower hyperfine field in the SPS misfit suggests that electronic spins ($S$=1-2) of the RS Co$^{3+}$ ions in this sample are not completely frozen even below 1 K. They seem to fluctuate slowly in a time scale comparable to relaxation times of nuclear spins.

The key issue of the specific heat in [Ca$_2$CoO$_{3-t}$]$_{0.62}$(CoO$_2$) misfits is, above all, the existence of a Schottky peak in the 1-10 K range that shifts with magnetic field to higher temperatures. Our interpretation is based on an analogy with FM perovskite cobaltites Nd$_{0.7}$Sr$_{0.3}$CoO$_3$, Nd$_{0.7}$Ca$_{0.3}$CoO$_3$ and partially also Pr$_{0.7}$Ca$_{0.3}$CoO$_3$, where similar peak occurs due to a presence of the Nd$^{3+}$ and Pr$^{4+}$ ions in Kramers doublet state [25]. Therefore, we relate the Schottky peak in present misfits to a formation in CoO$_2$ layers of relatively stable LS Co$^{4+}$ species with spin doublet state ($S$=½) that is split by internal and external magnetic fields. Based on this hypothesis we can obtain the concentration of such strongly localized holes $n_{LSCo4+}$ by a simple integration over the temperature range of Schottky peak, $n_{LSCo4+} = \int \frac{c_{Schottky}}{T} dT /(k_B N_{Av} \ln 2)$. Based on the $C_p/T$ data at zero external field (the baseline for Schottky peak is represented in Figs.10 and 11 by dotted lines), the concentration of strongly localized holes is determined to $n_{LSCo4+}$=0.054 per CoO$_2$ for the highly oxygenized sample CCR (total doping $n$=0.33 hole per CoO$_2$), and $n_{LSCo4+}$=0.031 per CoO$_2$ for the oxygen deficient sample SPS ($n$=0.17 hole per CoO$_2$). As the most obvious reason for such massive localization of hole carriers we consider the gain of Zeeman energy in staggered field $H_{rs}$ that is created by the AFM correlated IS or HS Co$^{3+}$ spins in adjacent RS blocks and is further enhanced under applied magnetic field. The location of the Schottky peak maximum $T_{max}$ in the $C_p/T$ data bears information on average staggered field $H_{rs}$ acting on the LS Co$^{4+}$ doublet $S$=½. Based on $T_{max}$ = 7.7 and 2.3 K and using the expression $T_{max} = 0.305 g\mu_B H_{rs}/k_B$ and $g \approx 2.2$, we get $H_{rs}$=170 kOe and 50 kOe for the CCR and SPS samples, respectively. (The *g*-factor value slightly above the free electron value was deduced earlier in the less magnetically complicated systems $Na_xCoO_2$ and bismuth based misfits [13, 14, 26], where $H_{rs}=0$ because of absence of RS spins.)

The analysis of Schottky peaks under external fields remains open. Heat capacity data in Figs. 10 and 11 evidence a rather fast shift of the peak toward higher temperatures, and suggest some increase of its integral value, but accurate determination is difficult for an uncertainty of the background, in which, namely, the $\gamma T$ term is both the field and temperature dependent. As a first approximation, we may just suppose that the orientations of staggered field are at random and the average effect is given by a "vector" sum of the staggered and external fields, $H = \sqrt{H_{rs}^2 + H_{ext}^2}$. Finally let us note that at the lowest temperature there will be also a contribution of the Weiss molecular field due to FM spin correlations in the $CoO_2$ sublattice, $H_m = \kappa <M_{CoO2}>$.

*Thermoelectric power*

Let us underline first that the above derived concentration of stable LS $Co^{4+}$ species represents practically the same section of the total carrier concentration, i.e. $n_{LSCo4+}/n=0.16-0.17$, for both $[Ca_2CoO_{3-t}]_{0.62}(CoO_2)$ misfits which differ very much in oxygen content. This proportionality with total doping level suggests that LS $Co^{4+}$ detected by the heat capacity experiments are not mere defects, but belong likely to an equilibrium distribution, which involves both the itinerant and more-or-less localized carrier states. We thus presume that the electronic transport in $[Ca_2CoO_{3-t}]_{0.62}(CoO_2)$ misfits is carried partly by the coherent motion and partly by the incoherent hopping.

The main task is then the theoretical interpretation of the magnetothermopower observed on the $O_2$ and Ar-annealed samples $[Ca_2CoO_{3-t}]_{0.62}(CoO_2)$, see Figs. 8 and 9. Our starting point is based on recent works of Shastry et al. and Zlatic et al. who demonstrated the applicability of Kelvin thermopower formula for layered cobaltates [11, 12]. We therefore follow this thermodynamic potential approach and neglect kinetic characteristics like *k*-space variations of Fermi velocities and energy variations of relaxation times. In the sense of "entropy per carrier", the Seebeck coefficient can be viewed as a superposition of an entropy part due to "metallic" (quasiparticle) carriers (see the expression A.2 in Appendix A) and the entropy part due to holes that jump in certain energy landscape of local sites and possess fast fluctuating spins. Latter part involves two contributions according to Appendix B – the first one is a formal analogy to A.2 and depends on the spectral distribution of spinless fermions (expression B.1), and the second one is due to bare spins and is closely related to standard formula for entropy of the spin ½ paramagnet (expression B.2).

At low enough temperatures, the theory gives a sum of two terms:

$$S_{Kelvin} = \frac{\pi^2 k_B^2 T}{3q_e}\left(\frac{\partial \ln N}{\partial \varepsilon}\right)_\mu + \frac{n_l}{n}\frac{k_B}{|q_e|}[\ln(2\cosh(g\mu_B H/2k_B T)) - (g\mu_B H/2k_B T)\tanh(g\mu_B H/2k_B T)], \quad (4)$$

where $n$ and $n_l$ denote the total and "localized" carrier concentration, respectively, $g$ stands for spin $g$ factor and $H$ comprises the total magnetic field affecting the charge carriers.

The first term in formula (4) is the $T$-linear thermopower with slope proportional to the logarithmic derivation of density of states $\frac{\partial \ln N}{\partial \varepsilon}$ at Fermi level $\mu=\mu_0$ (see expression A.3 in Appendix A). Such term is commonly known for metals, but when applied to a coexistence of the itinerant and localized states, it should comprise the overall density of states. The second term of (4) that depends on $g\mu_B H/2k_B T$ comes from the spin-entropy part and refers to more-or-less localized hole states that do not allow double occupation (spin-up and spin-down). It is thus presumed that these holes keep some individual spin-degree of freedom, which gives a positive addition to thermopower. The theoretical limit at zero field makes $S_{Kelvin,2} = \frac{k_B}{|q_e|}\ln 2 = 59$ µV/K, but the actual value will depend on the effective ratio of hopping to total charge carrier concentration, $n_l/n$.

The character of the thermopower data for present $[Ca_2CoO_{3-t}]_{0.62}(CoO_2)$ samples A and B, plotted in Figs. 8 and 9, confirms the consistency with above presented theoretical predictions. At the lowest accessible temperature of about 2 K, the magnetic field dependence of Seebeck coefficient shows a steep decrease already at small fields applied, and reaches the saturation at 70 kOe. The total Seebeck reduction associated with the spin-entropy contribution makes about 15-20 µV/K for both the $O_2$- annealed and Ar-annealed samples, respectively. As the temperature is raised, the initial slope of the field dependence is quickly decreased and the saturation is shifted to higher fields. The total change of Seebeck coefficient is increased to ≈30 µV/K for $T$>7 K. This value corresponds to 50% of the theoretical spin-entropy contribution $\frac{k_B}{|q_e|}\ln 2 = 59$ µV/K, from which we can deduce that $n_l/n$ ≈0.50 for both our samples.

The observed field and temperature dependences cannot be, however, simply explained by considering the spin entropy of $CoO_2$ holes as a function of $g\mu_B H/2k_B T$ (see the second term of expression (4)), taking into account only the external field, as it was done in previous magnetothermopower studies of related $Na_xCoO_2$ or bismuth based misfit [13, 14]. Actually there is a concurrence of two internal effects in the present case of $[Ca_2CoO_{3-t}]_{0.62}(CoO_2)$ misfits. First, there is the staggered field created by IS or HS $Co^{3+}$ spins in adjacent RS layers. As shown by the heat capacity data, the staggered field seen by stable LS $Co^{4+}$ may reach up to $H_{rs}$=170 kOe but, depending on hopping frequency of mobile holes, a motional averaging takes place, so we anticipate a broad span of non-zero local fields, which will reduce the spin entropy even at absence of external field. This may explain the decreased magnetothermopower observed at the lowest temperatures but cannot elucidate its initial steepness since increasing external field has only moderate effect according to "vector" summation $H = \sqrt{H_{rs}^2 + H_{ext}^2}$. And second, any attempt to analyse quantitatively the magnetothermopower must include also the effect of Weiss molecular field due to the short-range FM ordering of spins in the $CoO_2$ layers.

The last comment concerns the high-temperature behaviour of thermoelectric power in present misfits. At the high temperature limit, the spin entropy part of Seebeck coefficient in (4) is expected to remain on its zero-field value $S_{Kelvin,2} = \frac{n_l}{n} \frac{k_B}{|q_e|} \ln 2$, while the low-temperature $T$-linear part converge toward a saturation given by the modified Heikes thermopower

$$S_{Heikes} = \frac{k_B}{|q_e|} \left[ \ln \frac{1-n}{n} - \left(1 - \frac{n_l}{n}\right) \ln \left(1 - \frac{n_l}{n}\right) - \frac{n_l}{n} \ln \frac{n_l}{n} \right]. \qquad (5)$$

Here, the first term depending on the total doping $n$ reminds the common Heikes formula that derives from the entropy of the $(1-n)Co^{3+}/nCo^{4+}$ mixing, while the remaining terms dependent on $\frac{n_l}{n}$ are derived by supposing that the $n-n_l$ states allowing double hole occupation and the $n_l$ states allowing single hole occupation are distinguishable even at high temperatures. If $\frac{n_l}{n} = 0$ or 1, this excessive contribution to the mixing entropy part is identically zero, while it achieves a maximum value of $\frac{k_B}{|q_e|} \ln 2$ for $n_l/n=0.5$, which is actually our experimentally determined value. In particular, considering the $O_2$ annealed sample of total doping $n=0.31$ hole per $CoO_2$ and supposing that the ratio $n_l/n=0.5$ is not essentially temperature dependent, the calculation gives a high temperature estimate $S_{Kelvin,2} = 0.5 \frac{k_B}{|q_e|} \ln 2 = 30$ μV/K and $S_{Heikes} = \frac{k_B}{|q_e|} \left[ \ln \frac{1-0.31}{0.31} \right] + \frac{k_B}{|q_e|} \ln 2 = 125$ μV/K. This corresponds in total to the Seebeck coefficient of 155 μV/K, which is in a very close agreement with the experimentally observed data at high temperatures for this sample.

## V. Conclusions

The oxygenized and reduced samples of the $[Ca_2CoO_{3-t}]_{0.62}(CoO_2)$ misfit cobaltate were obtained by the $O_2$- and Ar-annealing of the as-prepared ceramic material. Large negative magnetoresistence is observed at low temperatures, while the phonon-dominated thermal conductivity does not show any perceptible impact of the applied field. Huge field effect is found in thermopower and the observed Seebeck coefficients are interpreted in terms of "entropy per carrier" following Kelvin formula $S_{Kelvin} = \frac{1}{q_e}\left(\frac{\partial S}{\partial n}\right)$. It is argued that apart from "metallic-like" carriers in the $CoO_2$ layers there is a large equilibrium fraction of holes in partially or fully localized states with little mutual interaction, contributing to entropy by their individual spin degrees of freedom. The number of localized holes $n_l$ is shown to increase linearly with the total hole doping $n$. In such a situation, the low-temperature thermopower can be expressed as a sum of the $T$-linear "metallic-like" part and the spin-entropy part that can be suppressed by magnetic field if temperature is low enough. The spin-entropy contribution actually observed is very similar for the $O_2$- and Ar-annealed samples, irrespective their very different doping levels. It makes 15-20 µV/K at $T$= 2 K and is increased to about 30 µV/K for $T$>7 K. This latter value corresponds to 50% of the theoretical limit $\frac{k_B}{|q_e|}\ln 2 = 59$ µV/K.

The spin-entropy contribution to thermopower of $[Ca_2CoO_{3-t}]_{0.62}(CoO_2)$ misfit is currently interpreted within an *ad hoc* model that is based on the notion of so-called Spin liquid. Although the model cannot describe the obviously very complex correlated character of carriers in the $CoO_2$ layers and limits just to the exclusion of double occupation at a given site, it seems to capture all essential features of observed thermopower. More realistic models of the correlations are still to be searched, which is still more demanding for the electrical conductivity, namely for understanding of its VRH characteristics at low temperatures, the result that has been found also on the $[Ca_2CoO_{3-t}]_{0.62}(CoO_2)$ single crystals and is probably of general validity [27]. An open question is also the anomalous behaviour of Hall coefficient that decreases with increasing temperature in a $1/T$ manner, exhibits a minimum at around 100 K and takes an increasing $T$-linear trend at the medium and high temperatures, see the data reported elsewhere [6, 8, 14, 28].

**Acknowledgments**

We acknowledge numerous discussions with J. Kuneš and P. Novák. The work was performed with the financial support of the Grant Agency of the Czech Republic within the Project No. GA13-17538S and the program of Czech Research Infrastructures (project no. LM2011025).

# Appendix A: Kelvin thermopower in Fermi gas or liquid

The thermopower phenomena are commonly treated in two distinct approaches. In the mixed-valence systems with hopping mechanism, Heikes formula is generally applied. It directly follows from Kelvin expression (see formula (A.1) below), taking into account the entropy of mixing. On the other hand, the generalized Mott formula (1) in the main text, is derived from Boltzmann transport equations, but with some precautions it can be formulated in terms of the entropy of mixing, as well. To illustrate the relationship between the Mott and Kelvin formulas, let us consider a system of non-interacting fermions and start with the grand potential $\Phi$, which is commonly calculated from partition function $Z_{gr}$ (see e. g. Ref. [29])

$$\Phi = U - TS - \mu n = -k_B T \ln(Z_{gr}),$$

where $U$ is the internal energy, $S = -\partial\Phi/\partial T$ is the entropy, and $n = -\partial\Phi/\partial\mu$ is the average number of fermions in the system, all considered per unit volume. In the case of quasicontinuous spectrum of microscopic states (band-like or localized eigenfuctions, characterized by the density of states $N(\varepsilon)$), the chemical potential $\mu$ is implicitly defined by the expression

$$n = \int_0^W \frac{N(\varepsilon)}{1+\exp((\varepsilon-\mu)/k_B T)} d(\varepsilon) \text{ or } n = \int_0^W N(\varepsilon) f(\varepsilon) d(\varepsilon),$$

when Fermi function $f = \dfrac{1}{1+\exp((\varepsilon-\mu)/k_B T)}$ is introduced.

The internal energy and entropy are given by

$$U = \int_0^W \frac{\varepsilon N(\varepsilon)}{1+\exp((\varepsilon-\mu)/k_B T)} d(\varepsilon)$$

$$S = \frac{1}{T}\int_0^W N(\varepsilon)\left(\frac{\varepsilon-\mu}{1+\exp((\varepsilon-\mu)/k_B T)} + k_B T \ln(\exp(-(\varepsilon-\mu)/k_B T)+1)\right) d(\varepsilon),$$

which can be rewritten, to show the relation to the entropy of mixing, as

$$S = -k_B \int_0^W N(\varepsilon)\left((1-f)\ln(1-f) + f\ln f\right) d(\varepsilon).$$

Seebeck coefficient is obtained using Kelvin formula [11]

$$S_{Kelvin} = \frac{1}{q_e}\left(\frac{\partial S}{\partial n}\right). \tag{A.1}$$

In our case,

$$S_{Kelvin} = \frac{1}{q_e}\left(\frac{\partial S}{\partial n}\right) = \frac{k_B}{q_e}\left(\frac{\partial \mu}{\partial n}\right)\int_0^W N(\varepsilon)\left(\ln\frac{1-f(\varepsilon)}{f(\varepsilon)}\right)\frac{\partial f(\varepsilon)}{\partial \mu}d(\varepsilon),$$

where $\left(\frac{\partial \mu}{\partial n}\right)_{H,T} = 1/\int_0^W N(\varepsilon)\frac{\partial f(\varepsilon)}{\partial \mu}d(\varepsilon)$, and

$$\ln\frac{1-f(\varepsilon)}{f(\varepsilon)} \equiv (\varepsilon - \mu)/k_B T,$$

$$\frac{\partial f(\varepsilon)}{\partial \mu} = -\frac{\partial f(\varepsilon)}{\partial \varepsilon} = \frac{\exp((\varepsilon-\mu)/k_B T)}{[1+\exp((\varepsilon-\mu)/k_B T)]^2} \equiv f(1-f).$$

From these expressions, the thermopower formula (3) in the main text directly follows:

$$S_{Kelvin} = \frac{k_B}{q_e}\int_0^W N(\varepsilon)\frac{(\varepsilon-\mu)}{k_B T}\left(-\frac{\partial f(\varepsilon)}{\partial \varepsilon}\right)d(\varepsilon)/\int_0^W N(\varepsilon)\left(-\frac{\partial f(\varepsilon)}{\partial \varepsilon}\right)d(\varepsilon) \qquad (A.2)$$

For $T \ll W/k_B$, the calculation of thermopower is manageable considering that $\left(-\frac{\partial f(\varepsilon)}{\partial \varepsilon}\right)$ in the integrands has a symmetrical peak at $\varepsilon = \mu$ and behaves like a delta-function of width $k_B T$ there. In that case the density of states can be expanded into a Taylor series about this point, $N(\varepsilon) = N(\mu) + \left(\frac{\partial N}{\partial \varepsilon}\right)_\mu (\varepsilon - \mu) + \ldots$.

The integrands can be then expressed as a sum of terms $(\varepsilon-\mu)^p\left(-\frac{\partial f(\varepsilon)}{\partial \varepsilon}\right)$ whose integration can be performed analytically. Thus, if temperature is low enough so that the interval $\langle \mu - k_B T, \mu + k_B T\rangle$ is well inside the bandwidth and $N(\varepsilon)$ has no sharp features there, the integrals of the $(\varepsilon-\mu)^p\left(-\frac{\partial f(\varepsilon)}{\partial \varepsilon}\right)$ type are zeros for $p$-odd and, for $p$-even, they can be calculated analytically (see e. g. [30]). In the first approximation, the thermopower acquires the $T$-linear form, familiar for metals:

$$S_{Kelvin} = \frac{1}{q_e T N(\mu)}\left(\frac{\partial N}{\partial \varepsilon}\right)_\mu \int_0^W (\varepsilon-\mu)^2 \frac{\exp((\varepsilon-\mu)/k_B T)}{[1+\exp((\varepsilon-\mu)/k_B T)]^2}d(\varepsilon) = \frac{\pi^2 k_B^2 T}{3q_e}\left(\frac{\partial \ln N}{\partial \varepsilon}\right)_\mu. \qquad (A.3)$$

This formula differs from common Mott thermopower by a neglect of the relaxation time $\tau$ and particle velocity $v_x$ in the logarithm, see [11].

It is important that formulas (A.2) and (A.3) are derived for ensemble of fermions with inclusion of their spin degeneracy, yet no extra spin-entropy contribution, as given by expression (3) in the main text, appears in Fermi liquid. One may, of course, consider the role of external magnetic field in the Zeeman splitting $\pm\mu_B H$ of the spin-down and spin-up states, but its effect on Seebeck coefficient is of a second order and thus small.

**Appendix B: Kelvin thermopower in Spin liquid model**

It is characteristic for Fermi quasiparticles that their band-like states allow a simultaneous spin-up and spin-down occupation. This need not to be true for the mixed-valence system with electron or hole carriers hopping over nearly localized states. Namely, any on-site coincidence of two carriers would be penalized due to Coulombic repulsion. The fermionic model in which the repulsion is so strong that double occupancy is forbidden was introduced by Spałek and his collaborators, and was named as Spin liquid [16, 17, 18].

The occupation of a particular level $\varepsilon$ with the spin-up or spin-down electron carrier is given by conditional probabilities

$$f_\uparrow(\varepsilon) = (1 - f_\downarrow(\varepsilon)) \frac{1}{1 + \exp((\varepsilon - \mu_B H - \varsigma)/k_B T)}$$

$$f_\downarrow(\varepsilon) = (1 - f_\uparrow(\varepsilon)) \frac{1}{1 + \exp((\varepsilon + \mu_B H - \varsigma)/k_B T)},$$

where $\varepsilon$ is the energy of the state in the absence of external field ($H=0$) [17]. From these two formulas, we get the total occupation regardless the carrier spin orientation

$$f(\varepsilon) = f_\uparrow(\varepsilon) + f_\downarrow(\varepsilon) = \frac{1}{1 + \frac{\exp((\varepsilon - \varsigma)/k_B T)}{2\cosh(\mu_B H/k_B T)}} \equiv \frac{1}{1 + \exp((\varepsilon - \mu)/k_B T)},$$

This distribution function is specific by the field-dependent prefactor $1/(2\cosh(\mu_B H/k_B T))$ or $1/2$ for $H=0$. In the last expression, this prefactor is included in the parameter $\mu = \varsigma + k_B T \ln(2\cosh(\mu_B H/k_B T))$. Let us note that in exact analogy to standard Fermi function it is the actual value $\mu$, subjected to the condition $\int_0^W N(\varepsilon) f(\varepsilon) d(\varepsilon) = n$, that defines the position of chemical potential in the energy spectrum of Spin liquid. In particular, both in the Fermi and Spin liquids, the chemical potential shifts as $T^2$ at low temperature, with the same quadratic coefficient dependent on the derivation of the density of states at Fermi level, $\left(\frac{\partial N}{\partial \varepsilon}\right)_\mu$. One should be only aware that the number of available states in Spin liquid is half of that for Fermi gas or liquid.

Another useful quantity is the difference of spin-up and spin-down probabilities

$$f_\uparrow(\varepsilon) - f_\downarrow(\varepsilon) = \tanh(\mu_B H/k_B T) \frac{1}{1 + \frac{\exp((\varepsilon - \varsigma)/k_B T)}{2\cosh(\mu_B H/k_B T)}},$$

from which it immediately follows the paramagnetic contribution of the $\varepsilon$-state to the total magnetization, in Bohr magnetons,

$$m_\varepsilon = \frac{f_\uparrow(\varepsilon) - f_\downarrow(\varepsilon)}{f_\uparrow(\varepsilon) + f_\downarrow(\varepsilon)} = \tanh(\mu_B H/k_B T).$$

This shows that the common $S=\frac{1}{2}$ formula for magnetic moment, $M = n\,\mu_B \tanh(\mu_B H/k_B T)$, is valid in Spin liquid for any dispersion of the $\varepsilon$-states and their occupation.

The internal energy $U$ for such ensemble of $n$ carriers is given by the formula:

$$U = \int_0^W [(\varepsilon - \mu_B H)f_\uparrow(\varepsilon) + (\varepsilon + \mu_B H)f_\downarrow(\varepsilon)]N(\varepsilon)d(\varepsilon) = U_1 + U_2,$$

where $U_1 = \int_0^W \varepsilon\,N(\varepsilon)f(\varepsilon)d(\varepsilon)$ and $U_2 = -n\mu_B H \tanh(\mu_B H/k_B T)$.

This formula for internal energy $U$ may serve e. g. for the calculation of electronic heat, see the original work of Spałek and Wójcik [16]. Our concern is, however, the calculation of Seebeck coefficient through the formula for entropy $S$:

$$S = -k_B \int_0^W [(1 - f_\uparrow(\varepsilon) - f_\downarrow(\varepsilon))\ln(1 - f_\uparrow(\varepsilon) - f_\downarrow(\varepsilon)) + f_\uparrow(\varepsilon)\ln f_\uparrow(\varepsilon) + f_\downarrow(\varepsilon)\ln f_\downarrow(\varepsilon)]N(\varepsilon)d(\varepsilon) = S_1 + S_2,$$

where

$$S_1 = -k_B \int_0^W [(1 - f(\varepsilon))\ln(1 - f(\varepsilon)) + f(\varepsilon)\ln f(\varepsilon)]N(\varepsilon)d(\varepsilon) \text{ and}$$

$$S_2 = -n\,k_B \left[\frac{1 - m_\varepsilon}{2}\ln\frac{1 - m_\varepsilon}{2} + \frac{1 + m_\varepsilon}{2}\ln\frac{1 + m_\varepsilon}{2}\right] =$$
$$-n\,k_B \left[\frac{1 - \tanh(\mu_B H/k_B T)}{2}\ln\frac{1 - \tanh(\mu_B H/k_B T)}{2} + \frac{1 + \tanh(\mu_B H/k_B T)}{2}\ln\frac{1 + \tanh(\mu_B H/k_B T)}{2}\right]$$

which can be rewritten to a more common form for the spin $\frac{1}{2}$ paramagnet

$$S_2 = n\,k_B [\ln(2\cosh(\mu_B H/k_B T)) - (\mu_B H/k_B T)\tanh(\mu_B H/k_B T)].$$

It is seen that both $U$ and $S$ can be separated into two terms corresponding to spinless fermions and isolated spins, respectively.

Seebeck coefficient can be expressed again as a sum of two terms. The first one depends on details of density of states and is expressed by a formula analogical to Mott thermopower:

$$S_{Kelvin,1} = \frac{1}{q_e}\left(\frac{\partial S_1}{\partial n}\right) = \frac{k_B}{q_e}\int_0^W N(\varepsilon)\frac{(\varepsilon - \mu)}{k_B T}\left(-\frac{\partial f(\varepsilon)}{\partial \varepsilon}\right)d(\varepsilon) / \int_0^W N(\varepsilon)\left(-\frac{\partial f(\varepsilon)}{\partial \varepsilon}\right)d(\varepsilon). \qquad \text{(B.1)}$$

The second term of Seebeck coefficient is purely due to spin entropy and its calculation for a paramagnet is trivial

$$S_{Kelvin,2} = \frac{1}{|q_e|}\left(\frac{\partial S_2}{\partial n}\right) = \frac{k_B}{|q_e|}[\ln(2\cosh(\mu_B H/k_B T)) - (\mu_B H/k_B T)\tanh(\mu_B H/k_B T)] \quad (B.2)$$

For scaling purposes, we note that the $S_{Kelvin,2}$ term depends on magnetic field through the variable $h = \mu_B H/k_B T$, where $H$ contains not only the external field but also, in the case of spontaneous magnetic ordering, the local molecular field. For $H$=0 or for $k_B T \gg \mu_B H$, $S_{Kelvin,2} = \frac{k_B}{|q_e|}\ln 2 = 59$ μV/K.